\documentclass[a4paper,11pt]{article}
\usepackage{amssymb}
\usepackage{graphicx}
\usepackage{amsmath}

 \DeclareFontFamily{U}{rsf}{}
\DeclareFontShape{U}{rsf}{m}{n}{
  <5> <6> rsfs5 <7> <8> <9> rsfs7 <10-> rsfs10}{}
\DeclareMathAlphabet\Scr{U}{rsf}{m}{n} \makeatletter
\@addtoreset{equation}{section} \makeatother

\def\be{\begin{equation}}
\def\ee{\end{equation}}
\def\ba{\begin{array}}
\def\ea{\end{array}}

\newcommand{\bea}{\begin{eqnarray}}
\newcommand{\eea}{\end{eqnarray}}

\begin{document}

\begin{titlepage}
 \thispagestyle{empty}
\begin{flushright}
     \hfill{CERN-PH-TH/2011-006}\\
 \end{flushright}

 \vspace{50pt}

 \begin{center}
     { \huge{\bf      {Two-Centered Magical Charge Orbits}}}

     \vspace{25pt}

     {\Large {Laura Andrianopoli$^{a}$, Riccardo D'Auria$^{a}$, Sergio Ferrara$^{b,c}$, \\\vspace{2pt} Alessio Marrani$^{b}$ and Mario Trigiante$^{a}$}}

     \vspace{20pt}

     {\it ${}^a$ Dipartimento di Fisica, Politecnico di Torino,\\
     Corso Duca degli Abruzzi 24, I-10129, Italy,\\
     and INFN, Sezione di Torino, Italy;\\
     \texttt{laura.andrianopoli@polito.it}\\
     \texttt{riccardo.dauria@polito.it}\\
     \texttt{mario.trigiante@polito.it}}

     \vspace{10pt}

  {\it ${}^b$ Physics Department, Theory Unit, CERN,\\
     CH -1211, Geneva 23, Switzerland;\\
     \texttt{sergio.ferrara@cern.ch}\\
     \texttt{alessio.marrani@cern.ch}}

     \vspace{10pt}

    {\it ${}^c$ INFN - Laboratori Nazionali di Frascati,\\
     Via Enrico Fermi 40, I-00044 Frascati, Italy}

     \vspace{100pt}

     {ABSTRACT}

 \vspace{10pt}
 \end{center}
We determine the two-centered generic charge orbits of magical
$\mathcal{N}=2 $ and maximal $\mathcal{N}=8$ supergravity theories
in four dimensions. These orbits are classified by seven $U$-duality
invariant polynomials, which group together into four invariants
under the horizontal symmetry group $SL\left( 2,\mathbb{R}\right) $.
These latter are expected to disentangle different physical
properties of the two-centered black-hole system. The invariant with
the lowest degree in charges is the symplectic product $\left\langle \mathcal{Q}_{1},\mathcal{Q%
}_{2}\right\rangle $, known to control the mutual non-locality of
the two centers.
\end{titlepage}

\section{\label{Intro}Introduction}

Multi-centered black-hole solutions of supergravity theories in $d=4$
space-time dimensions have recently received much attention, especially in
connection to the classification of non-perturbative string BPS states and
their brane interpretation \cite{DM-1}. A generalisation of the \textit{%
attractor Mechanism} \cite{AM-Refs,FGK} (for a review, see \textit{e.g.}
\cite{ADFT-rev}) has been shown to occur, as firstly pointed out by Denef
\cite{D-1}, called split attractor flow for BPS $\mathcal{N}=2$ black holes
\cite{D-1,D-2,BD,DM-1}.

Attempts to generally classify the two-centered solutions of supergravity
theories with symmetric scalar manifolds and electric-magnetic duality ($U$%
-duality\footnote{%
Here $U$-duality is referred to as the ``continuous'' symmetries of \cite
{CJ-1}. Their discrete versions are the $U$-duality non-perturbative string
theory symmetries introduced by Hull and Townsend \cite{HT-1}.}) symmetry
given by classical Lie groups have been considered \cite
{David,MS-FMO-1,FMOSY-1}. In particular, within the framework of the \textit{%
minimal coupling} \cite{Luciani} of vector multiplets to $\mathcal{N}=2$
supergravity, it was shown in \cite{MS-FMO-1} that different physical
properties, such as marginal stability and split attractor flow solutions,
can be classified by duality-invariant constraints, which in this case
involve two dyonic black-hole charge vectors, and not only one.

This leads one to consider the mathematical issue of the classification of
orbits of two (or more) dyonic charge vectors in the context of
multi-centered black-hole physics. For the theories treated in \cite
{MS-FMO-1,FMOSY-1}, the charge vector lies in the fundamental representation
of $U\left( 1,n\right) $ (\textit{minimally coupled} $\mathcal{N}=2$
supergravity \cite{Luciani}) and in the spinor-vector representation of $%
SL\left( 2,\mathbb{R}\right) \times SO\left( q,n\right) $, corresponding to
reducible cubic $\mathcal{N}=2$ sequence \cite{GST,dWVVP} for $q=2$, and to
matter-coupled $\mathcal{N}=4$ supergravity for $q=6$.

In \cite{MS-FMO-1}, the two-centered $U$-invariant polynomials of the
\textit{minimally coupled} theory were constructed, and shown to be four
(dimension of the adjoint of the two-centered horizontal symmetry $U\left(
2\right) $). The same was done for the aforementioned cubic sequence in \cite
{FMOSY-1}, where the number of $U$-invariants were computed to be seven for $%
n\geqslant 2$, six for $n=1$ and five for the irreducible $t^{3}$
model.\smallskip

It is the aim of the present investigation to generalise these results to
four-dimensional supergravity theories with symmetric \textit{irreducible}
scalar manifolds, in particular to the $\mathcal{N}=8$ maximal theory and to
the $\mathcal{N}=2$ magical models.

We find that when the stabilizer of a two-centered charge orbit is \textit{%
non-compact}, the corresponding orbit is \textit{not} unique. As we will
consider in Section \ref{Non-Compact}, this feature is also exhibited by the
classification of the orbits of two non-lightlike vectors in a
pseudo-Euclidean space $E_{p,q}$ of dimension $p+q$ and signature $\left(
p,q\right) $. A prominent role is played by an \textit{emergent} horizontal
symmetry $SL_{h}\left( 2,\mathbb{R}\right) $, whose invariants classify all
possible two-vector orbits.\smallskip

In this respect, the aforementioned $t^{3}$ model, whose $U$-duality group
is $SL\left( 2,\mathbb{R}\right) $, provides a simple yet interesting
example, because it may be obtained both as rank-$1$ truncation of the
\textit{reducible} symmetric models and as first, non-generic element of the
sequence of \textit{irreducible} $\mathcal{N}=2$ symmetric models, which
contains the four rank-$3$ magical supergravity theories mentioned above.
The two-centered configurations and the generic (BPS) orbit $\mathcal{O}%
=SL\left( 2,\mathbb{R}\right) $ of $t^{3}$ model were studied in Sec. 7 of
\cite{FMOSY-1}, in which it was pointed out that, as it occurs also for the
one-centered case \cite{BFGM1}, no stabilizer for the two-centered orbit
exists\footnote{%
As it holds for the magical $J_{3}^{\mathbb{R}}$ model, see Table I.}. The
five components of the spin $s=2$ horizontal tensor $\mathbf{I}_{abcd}$
(defined in (\ref{I-def}) below, and explicitly given by (\ref{I+2})-(\ref
{I-2})) form a complete basis of duality-invariant polynomials \cite{FMOSY-1}%
; as a consequence, the counting (\ref{conta}) for $p=2$-centered black hole
solutions in the $t^{3}$ model simply reads $5+3-0=4\times 2$, because $%
I_{p=2}=5$ and dim$_{\mathbb{R}}(\mathcal{G}_{p})=0$. Moreover, there exist
only two independent $\left[ SL_{h}\left( 2,\mathbb{R}\right) \times
SL\left( 2,\mathbb{R}\right) \right] $-invariant polynomials, which can be
taken to be the symplectic product $\mathcal{W}$ (of order two in charges,
defined in (\ref{4}) below) and $\mathbf{I}_{6}$ (of order six in charges,
defined in (\ref{I6}) below); an alternative choice of basis for the $%
SL\left( 2,\mathbb{R}\right) $-invariant polynomials is thus \textit{e.g.}
given by three components of $\mathbf{I}_{abcd}$ out of the five (\ref{I+2}%
)-(\ref{I-2}), and the two horizontal invariants $\mathcal{W}$ and $\mathbf{I%
}_{6}$.

\newpage The plan of the paper is as follows.

In Section \ref{Counting-Analysis} we give a group theoretical method (based
on progressive branchings of symmetry groups, considered as complex groups)
to find the multi-centered charge orbits of a theory with a symmetric scalar
manifold; we then apply it to all \textit{irreducible} symmetric cases. The
analysis of this section will not depend on the real form of the stabilizer
of the orbit, and the results will then hold both for BPS and all the
non-BPS orbits of the given model. In Section \ref{Formal-Analysis} we
propose a complete basis for $U$-duality polynomials in the presence of two
dyonic black-hole charge vectors in irreducible symmetric models, and we
also consider the role of the horizontal symmetry in this framework. Section
\ref{Non-Compact} extends the analysis of Section \ref{Counting-Analysis} to
different non-compact real forms of the stabilizer of one-centered charge
orbits related to Jordan algebras over the octonions, namely to $\mathcal{N}%
=8$ theory (whose $\frac{1}{8}$-BPS one-centered stabilizer is $E_{6\left(
2\right) }$) and for exceptional magical $\mathcal{N}=2$ theory (whose BPS
and non-BPS $\mathcal{I}_{4}>0$ one-centered stabilizers are the compact $%
E_{6\left( -78\right) }$ and the non-compact $E_{6\left( -14\right) }$,
respectively).\smallskip

Possible extensions of the present investigation may also cover composite
configurations with ``small'' constituents, as well as a detailed study of
the multi-centered charge orbits in $\mathcal{N}=5,6$-extended supergravity
theories.

\section{\label{Counting-Analysis}Little Group of $p$ Charge Vectors in
Irreducible Symmetric Models}

We consider a $p$-center black hole solution in a Maxwell-Einstein
supergravity theory in $d=4$ space-time dimensions.

The $p$ dyonic black-hole charge vectors can be arranged as
\begin{equation}
\mathbf{Q}_{a}\equiv \left\{ \mathcal{Q}_{a}^{M}\right\} _{M=1,...,f},
\end{equation}
where $\mathcal{Q}_{a}^{M}$ sits in the irreducible representation $\left(
\mathbf{p},\mathbf{Sympl}\left( G_{4}\right) \right) $ of the group $%
SL_{h}\left( p,\mathbb{R}\right) \times G_{4}$. $\mathbf{p}$ is the
fundamental representation (spanned by the index $a=1,\dots ,p$) of the
horizontal symmetry group \cite{FMOSY-1} $SL_{h}\left( p,\mathbb{R}\right) $
(see Section \ref{Formal-Analysis}), while $\mathbf{Sympl}\left(
G_{4}\right) $ is the symplectic irreducible representation of the
black-hole charges, spanned by the index $M=1,\dots ,f$ of the $U$-duality
group $G_{4}$, where $f\equiv $dim$_{\mathbb{R}}\left( \mathbf{Sympl}\left(
G_{4}\right) \right) $.

Suppose there are $I_{p}$ independent $G_{4}$-invariant polynomials
constructed out of $\mathbf{Q}_{a}$, and let $\mathcal{G}_{p}$ denote the
little group of the system of charges, defined as the largest subgroup of $%
G_{4}$ such that $\mathcal{G}_{p}\,\mathbf{Q}_{a}=\mathbf{Q}_{a}$ $\forall a$%
. Then, the following relation\footnote{%
A necessary but not sufficient condition for Eq. (\ref{conta}) to hold is $%
p<f$, such that the $p$ dyonic charge vectors can all be taken to be
linearly independent.} holds \cite{FMOSY-1}:
\begin{equation}
I_{p}+\text{dim}_{\mathbb{R}}(G_{4})-\text{dim}_{\mathbb{R}}(\mathcal{G}%
_{p})=\,f\,p.  \label{conta}
\end{equation}

Some preliminary general observations are in order:

\begin{itemize}
\item  The group theoretical analysis of the present Section does not depend
on the real form of $G_{4}$ and $\mathcal{G}_{p}$. We will then generally
consider the complex groups. From a physical point of view, the BPS and
non-BPS cases in various supergravity theories correspond to different
choices of non-compact real forms of $\mathcal{G}_{p}$ (and of $G_{4}$, as
well). However, for BPS orbits in $\mathcal{N}=2$ symmetric models, and in
particular for magical models, the stabilizer is always the compact form of
the relevant group (see Table 1).

\item  We shall generally assume $\mathbf{Q}_{1}$ to be in a representation
corresponding to a ``large'' black hole\footnote{%
Multi-center configurations with ``small'' constituents \cite{BD,GLS-2,CS}
can be treated as well, and they will be considered elsewhere.}, namely such
that the quartic invariant $\mathcal{I}_{4}\left( \mathcal{Q}_{1}^{4}\right)
\neq 0$.

\item  We shall consider ``generic'' orbits, in which all $I_{p}$ invariants
are independent.

\item  There are two relevant cases, corresponding to different behaviors in
the counting of invariants:

\begin{enumerate}
\item[a)]  The largest subgroup commuting with $\mathcal{G}_{p}$ inside $%
G_{4}$ is $U(1)\subset G_{4}$, so that $\mathcal{G}_{p}\times U(1)\subset G$.

\item[b)]  A $U(1)$ commuting with $\mathcal{G}_{p}$ inside $G_{4}$ does not
exist.
\end{enumerate}

In the case $b$), all the singlets in the decomposition of $G_{4}\rightarrow
\mathcal{G}_{p}$ correspond to $p$-center $G_{4}$-invariant polynomials of $%
\mathbf{Sympl}\left( G_{4}\right) $. On the other hand, in the case $a$) the
number of singlets corresponds to the number of $p$-center $G_{4}$-invariant
polynomials, plus one if some of them are charged with respect to $U(1)$,
because one of the singlets can still be acted on by the corresponding $U(1)$%
-grading.

\item  The general method for working out $\mathcal{G}_{p}$ and thus $I_{p}$%
, having solved the problem for $p-1$ centers, is to consider the $p^{\text{%
th}}$ charge vector $\mathbf{Q}_{p}$ as transforming in a (reducible)
representation of the little group $\mathcal{G}_{p-1}$ of the former $p-1$
charges, and solve the corresponding one-charge-vector problem.
\end{itemize}

In the next Subsections we will consider the cases $p=1$ and $p=2$ in all
irreducible symmetric cases pertaining to supergravity theories in $d=4$
dimensions (with the exception of the rank-1 $t^{3}$ model, treated in \cite
{FMOSY-1}). In the case $p=1$, we will retrieve the well known result $%
I_{p=1}=1$, whereas in the $p=2$ case we will obtain $I_{p=2}=7$ in all
cases under consideration.

\begin{table}[h]
\begin{center}
\begin{tabular}{|c||c|}
\hline
$
\begin{array}{c}
\\
J_{3}^{\mathbb{A}}
\end{array}
$ & $
\begin{array}{c}
\\
\mathcal{O}_{p=2,BPS}=\frac{Conf\left( J_{3}^{\mathbb{A}}\right) }{\mathcal{G%
}_{p=2}\left( J_{3}^{\mathbb{A}}\right) } \\
~~
\end{array}
$ \\ \hline
$
\begin{array}{c}
\\
J_{3}^{\mathbb{O}} \\
~
\end{array}
$ & $\frac{E_{7\left( -25\right) }}{SO\left( 8\right) }$ \\ \hline
$
\begin{array}{c}
\\
J_{3}^{\mathbb{H}} \\
~
\end{array}
$ & $\frac{SO^{\ast }(12)}{\left[ SU\left( 2\right) \right] ^{3}}$ \\ \hline
$
\begin{array}{c}
\\
J_{3}^{\mathbb{C}} \\
~
\end{array}
$ & $\frac{SU\left( 3,3\right) }{\left[ U(1)\right] ^{2}}$ \\ \hline
$
\begin{array}{c}
\\
J_{3}^{\mathbb{R}} \\
~
\end{array}
$ & $Sp\left( 6,\mathbb{R}\right) $ \\ \hline
\end{tabular}
\end{center}
\caption{BPS generic charge orbits of $2$-centered extremal black holes in $%
\mathcal{N}=2$, $d=4$ magical models. $Conf\left( J_{3}^{\mathbb{A}}\right) $
denotes the ``conformal'' group of $J_{3}^{\mathbb{A}}$ (see \textit{e.g.}
\protect\cite{Gunaydin-Lects}, and Refs. therein). By introducing $\mathbb{A}%
=\mathbb{R}$, $\mathbb{C}$, $\mathbb{H}$, $\mathbb{O}$, it is worth
remarking that the stabilizer group $\mathcal{G}_{p=2}\left( J_{3}^{\mathbb{A%
}}\right) $ and the automorphism group $Aut\left( \mathbf{t}\left( \mathbb{A}%
\right) \right) $ of the \textit{normed triality} $\mathbf{t}\left( \mathbb{A%
}\right) $ in dimension dim$_{\mathbb{R}}\mathbb{A}=1$, $2$, $4$, $8$ (given
\textit{e.g.} in Eq. (5) of \protect\cite{Baez-O}) share the same Lie
algebra. In other words, $\frak{g}_{p=2}\left( J_{3}^{\mathbb{A}}\right)
\sim \frak{tri}\left( \mathbb{A}\right) $, where $\frak{tri}\left( \mathbb{A}%
\right) $ denotes the Lie algebra of $Aut\left( \mathbf{t}\left( \mathbb{A}%
\right) \right) $ itself (see \textit{e.g.} Eq. (21) of \protect\cite{Baez-O}%
). }
\end{table}

\subsection{\label{J3-O}$J_{3}^{\mathbb{O}}$ ($\mathcal{N}=2$), $J_{3}^{%
\mathbb{O}_{s}}$ ($\mathcal{N}=8$)}

Let us start considering the exceptional case, based on the Euclidean degree-%
$3$ Jordan algebra $J_{3}^{\mathbb{O}}$ on the octonions $\mathbb{O}$.
Since, as mentioned earlier, we actually work with complex groups, this case
pertains also to maximal $\mathcal{N}=8$ supergravity, based on the
Euclidean degree-$3$ Jordan algebra $J_{3}^{\mathbb{O}_{s}}$ on the split
octonions $\mathbb{O}_{s}$.

In the complex field, $G_{4}=E_{7}$ and $\mathbf{Sympl}\left( E_{7}\right) =%
\mathbf{Fund}\left( E_{7}\right) =\mathbf{56}$.

\begin{itemize}
\item  Let us first solve the one-center problem ($p=1$). $\mathcal{G}_{1}$
is a real form of $E_{6}$; the $\mathbf{56}$ branches with respect to $E_{6}$
as follows (subscripts denote the $U\left( 1\right) $-charges throughout):
\begin{equation}
\mathbf{56}\rightarrow \mathbf{1}_{-3}+\mathbf{27}_{-1}+\overline{\mathbf{27}%
}_{+1}+\mathbf{1}_{+3}\,,  \label{branche7}
\end{equation}
and correspondingly the charge vector $\mathbf{Q}_{1}$ (defined as $\left(
p^{\Lambda },q_{\Lambda }\right) $ throughout) decomposes as follows:
\begin{equation}
\mathbf{Q}_{1}=(p^{0},\mathbf{p}_{\mathbf{27}},q_{0},\mathbf{q}_{\overline{%
\mathbf{27}}})\,.
\end{equation}
Note that the branching (\ref{branche7}) contains two $E_{6}$-singlets, and $%
E_{7}\supset E_{6}\times U(1)=\mathcal{G}_{1}\times U(1)$. According to the
previous discussion, one of the singlets can be freely acted on by the $U(1)$%
. Thus, by acting with $G_{4}/\mathcal{G}_{1}=E_{7}/E_{6}$, the $1$-center
charge vector $\mathbf{Q}_{1}$ can be reduced as follows:
\begin{equation}
\mathbf{Q}_{1}\overset{E_{7}/E_{6}}{\longrightarrow }(I^{(1)},\mathbf{0}_{%
\mathbf{27}},\pm I^{(1)},\mathbf{0}_{\overline{\mathbf{27}}})\,.
\end{equation}
One is then left with only one independent singlet charge $I^{(1)}$ related
to the $1$-center quartic invariant $\mathcal{I}_{4}\left( \mathcal{Q}%
_{1}^{4}\right) $; therefore, $I_{1}=1$, as expected. This analysis is
consistent with the general formula (\ref{conta}), which in this case reads:
\begin{equation}
I_{1}+\text{dim}_{\mathbb{R}}(E_{7})-\text{dim}_{\mathbb{R}%
}(E_{6})=1+133-78=56\,.
\end{equation}

\item  Let us now proceed to deal with the two charge-vector problem ($p=2$%
). The second charge vector is denoted as $\mathbf{Q}_{2}\equiv (m^{\Lambda
},e_{\Lambda })$ throughout. Having solved the problem for $p=1$, we can
decompose $\mathbf{Q}_{2}$ with respect to $\mathcal{G}_{1}=E_{6}$ using (%
\ref{branche7}), obtaining the decomposition
\begin{equation}
\mathbf{Q}_{2}=(I^{(2)},\mathbf{m}_{\mathbf{27}},I^{(3)},\mathbf{e}_{%
\overline{\mathbf{27}}}),
\end{equation}
and then determine the corresponding little group inside $E_{6}$. The little
group of the irreducible representation $\mathbf{27}$ of $E_{6}$ is $F_{4}$,
under which
\begin{equation}
\mathbf{27}\rightarrow \mathbf{1}+\mathbf{26}\,,
\end{equation}
and correspondingly
\begin{equation}
\mathbf{m}_{\mathbf{27}}\rightarrow (I^{(4)},\mathbf{m}_{\mathbf{26}});\quad
\mathbf{e}_{\overline{\mathbf{27}}}\rightarrow (I^{(5)},\mathbf{e}_{\mathbf{%
26}}).
\end{equation}
Note in particular that $F_{4}$ is a maximal (symmetric) subgroup of $E_{6}$%
, so that all singlets correspond to extra $E_{7}$-invariant polynomials,
and that $\mathbf{m}_{\mathbf{26}}$ can be set to zero through the action of
$\mathcal{G}_{1}/F_{4}=E_{6}/F_{4}$, thus yielding the result:
\begin{equation}
\mathbf{Q}_{2}\overset{E_{6}/F_{4}}{\longrightarrow }(I^{(2)},I^{(4)},%
\mathbf{0}_{\mathbf{26}},I^{(3)},I^{(5)},\mathbf{e}_{\mathbf{26}}).
\end{equation}

\item  The $\mathbf{26}$ of $F_{4}$ has little group $SO(8)$, which does not
commute with a $U(1)$ in $F_{4}$. Under this non-maximal embedding, the $%
\mathbf{26}$ branches as
\begin{equation}
\mathbf{26}\rightarrow \mathbf{1}+\mathbf{1}+\mathbf{8_{v}}+\mathbf{8_{s}}+%
\mathbf{8_{c}}\,,
\end{equation}
and correspondingly
\begin{equation}
\mathbf{e}_{\mathbf{26}}\rightarrow (I^{(6)},I^{(7)},\mathbf{e_{8_{v}}},%
\mathbf{e_{8_{s}}},\mathbf{e_{8_{c}}})\,.
\end{equation}
Therefore, by acting with $F_{4}/\mathcal{G}_{2}=F_{4}/SO(8)$, $\mathbf{Q}%
_{2}$ can then be put in the form
\begin{equation}
\mathbf{Q}_{2}\overset{F_{4}/SO(8)}{\longrightarrow }(I^{(2)},I^{(4)},%
\mathbf{0}_{\mathbf{26}},I^{(3)},I^{(5)},I^{(6)},I^{(7)},\mathbf{0_{8_{v}}},%
\mathbf{0_{8_{s}}},\mathbf{0_{8_{c}}}).
\end{equation}
\end{itemize}

In conclusion, we found that the little group of a $2$-centered black-hole
solution is $\mathcal{G}_{2}=SO(8)$, and the corresponding $2$-centered
charge orbits correspond to different real forms of the quotient of complex
groups
\begin{equation}
\mathcal{O}_{p=2}=\frac{G_{4}}{\mathcal{G}_{2}}=\frac{E_{7}}{SO\left(
8\right) }.
\end{equation}
The $E_{7}$-invariant polynomials for a $2$-centered configuration are
seven: $I_{2}=7$; indeed, the general formula (\ref{conta}) gives:
\begin{equation}
I_{2}+\text{dim}_{\mathbb{R}}(E_{7})-\text{dim}_{\mathbb{R}%
}(SO(8))=7+133-28=112=2\cdot 56.
\end{equation}

\subsection{\label{J3-H}$J_{3}^{\mathbb{H}}$ ($\mathcal{N}=2\leftrightarrow
\mathcal{N}=6$)}

This model is based on the Euclidean degree-$3$ Jordan algebra $J_{3}^{%
\mathbb{H}}$ on the quaternions $\mathbb{H}$, and it is ``dual'' to $%
\mathcal{N}=6$ ``pure'' theory, because these theories share the same
bosonic sector \cite{GST,ADF,FGimK,ADFGT,RS}.

In the complex field $G_{4}=SO\left( 12\right) $, and $\mathbf{Sympl}\left(
SO\left( 12\right) \right) =\mathbf{32}$, the chiral spinor irreducible
representation of $SO\left( 12\right) $.

\begin{itemize}
\item  Let us first solve the problem for $p=1$. $\mathcal{G}_{1}$ is a real
form of $SU(6)$, the relevant (maximal symmetric) embedding is
\begin{equation}
SO(12)\supset SU(6)\times U(1)=\mathcal{G}_{1}\times U(1),  \label{emb-1}
\end{equation}
and the $\mathbf{32}$ accordingly branches
\begin{equation}
\mathbf{32}\rightarrow \mathbf{1}_{-3}+\mathbf{15}_{-1}+\overline{\mathbf{15}%
}_{+1}+\mathbf{1}_{+3}\,,  \label{branchso12}
\end{equation}
corresponding to the charge decomposition
\begin{equation}
\mathbf{Q}_{1}=(p^{0},\mathbf{p}_{\mathbf{15}},q_{0},\mathbf{q}_{\mathbf{15}%
})\,.
\end{equation}
The analysis here is completely analogous to the exceptional case above. The
branching (\ref{branchso12}) contains two $SU(6)$-singlets, but, by virtue
of (\ref{emb-1}), one of the singlets can be freely acted on by the $U(1)$.
By acting with $G_{4}/\mathcal{G}_{1}=SO\left( 12\right) /SU(6)$, $\mathbf{Q}%
_{1}$ can be reduced to
\begin{equation}
\mathbf{Q}_{1}\overset{SO\left( 12\right) /SU(6)}{\longrightarrow }(I^{(1)},%
\mathbf{0}_{\mathbf{15}},\pm I^{(1)},\mathbf{0}_{\mathbf{15}})\,,
\end{equation}
so that $I_{1}=1$, corresponding to the $1$-center quartic invariant $%
\mathcal{I}_{4}\left( \mathcal{Q}_{1}^{4}\right) $ only. Indeed, the general
formula (\ref{conta}) yields
\begin{equation}
I_{1}+\text{dim}_{\mathbb{R}}(SO(12))-\text{dim}_{\mathbb{R}%
}(SU(6))=1+66-35=32\,.
\end{equation}

\item  Let us consider now the $2$-centered case ($p=2$). Having solved the
problem for $p=1$, we further decompose $\mathbf{Q}_{2}$ with respect to $%
\mathcal{G}_{1}=SU(6)$:
\begin{equation}
\mathbf{Q}_{2}=\left( I^{(2)},\mathbf{m}_{\mathbf{15}},I^{(3)},\mathbf{e}_{%
\mathbf{15}}\right) ,
\end{equation}
and find the corresponding little group. The little group of the $\mathbf{15}
$ of $SU(6)$ is $USp\left( 6\right) $, under which such a representation
branches as follows:
\begin{equation}
\mathbf{15}\longrightarrow \mathbf{1}+\mathbf{14},
\end{equation}
yielding the charge decompositions
\begin{equation}
\mathbf{m}_{\mathbf{15}}\longrightarrow \left( I^{(4)},\mathbf{m}_{\mathbf{14%
}}\right) ;~~\mathbf{e}_{\mathbf{15}}\longrightarrow \left( I^{(5)},\mathbf{e%
}_{\mathbf{14}}\right) .
\end{equation}
Since $USp\left( 6\right) $ is maximally (and symmetrically) embedded in $%
SU\left( 6\right) $, all singlets correspond to extra $SO\left( 12\right) $%
-invariant polynomials, and $\mathbf{m}_{\mathbf{14}}$ can be set to zero
through the action of $\mathcal{G}_{1}/USp\left( 6\right) =SU\left( 6\right)
/USp(6)$, thus yielding the result:
\begin{equation}
\mathbf{Q}_{2}\overset{SU\left( 6\right) /USp(6)}{\longrightarrow }\left(
I^{(2)},I^{(4)},\mathbf{0}_{\mathbf{14}},I^{(3)},I^{(5)},\mathbf{e}_{\mathbf{%
14}}\right) .
\end{equation}

\item  The $\mathbf{14}$ (rank-$2$\ antisymmetric) of $USp(6)$ has little
group $\left[ SU\left( 2\right) \right] ^{3}$, which does not commute with a
$U\left( 1\right) $ in $USp(6)$. The $\mathbf{14}$ correspondingly branches
as
\begin{equation}
\mathbf{14}\longrightarrow \left( \mathbf{1},\mathbf{1},\mathbf{1}\right)
+\left( \mathbf{1},\mathbf{1},\mathbf{1}\right) +\left( \mathbf{1},\mathbf{2}%
,\mathbf{2}\right) +\left( \mathbf{2},\mathbf{2},\mathbf{2}\right) ,
\end{equation}
and thus
\begin{equation}
\mathbf{e}_{\mathbf{14}}\longrightarrow \left( I^{(6)},I^{(7)},\mathbf{e}%
_{\left( \mathbf{1},\mathbf{2},\mathbf{2}\right) },\mathbf{e}_{\left(
\mathbf{2},\mathbf{2},\mathbf{2}\right) }\right) .
\end{equation}
Therefore, by acting with $USp(6)/\mathcal{G}_{2}=USp(6)/$ $\left[ SU\left(
2\right) \right] ^{3}$, $\mathbf{Q}_{2}$ can then be put in the form
\begin{equation}
\mathbf{Q}_{2}\overset{USp(6)/\left[ SU\left( 2\right) \right] ^{3}}{%
\longrightarrow }(I^{(2)},I^{(4)},\mathbf{0}_{\mathbf{14}%
},I^{(3)},I^{(5)},I^{(6)},I^{(7)},\mathbf{0}_{\left( \mathbf{1},\mathbf{2},%
\mathbf{2}\right) },\mathbf{0}_{\left( \mathbf{2},\mathbf{2},\mathbf{2}%
\right) }).
\end{equation}
\end{itemize}

In conclusion, we found that the little group of a $2$-centered black-hole
solution is $\mathcal{G}_{2}=\left[ SU\left( 2\right) \right] ^{3}$, and the
corresponding $2$-centered charge orbit reads (in complexified form)
\begin{equation}
\mathcal{O}_{p=2}=\frac{G_{4}}{\mathcal{G}_{2}}=\frac{SO\left( 12\right) }{%
\left[ SU\left( 2\right) \right] ^{3}}.
\end{equation}
The $SO\left( 12\right) $-invariant polynomials for a $2$-centered
configuration are seven: $I_{2}=7$; indeed, the general formula (\ref{conta}%
) gives:
\begin{equation}
I_{2}+\text{dim}_{\mathbb{R}}(SO\left( 12\right) )-\text{dim}_{\mathbb{R}}(%
\left[ SU\left( 2\right) \right] ^{3})=7+66-9=64=2\cdot 32.
\end{equation}

\subsection{\label{J3-C}$J_{3}^{\mathbb{C}}$ ($\mathcal{N}=2$), $%
M_{1,2}\left( \mathbb{O}\right) $ ($\mathcal{N}=5$)}

Let us now consider the model based on the Euclidean degree-$3$ Jordan
algebra $J_{3}^{\mathbb{C}}$ on $\mathbb{C}$. Since, as mentioned earlier,
we actually deal with groups on the complex field, this case pertains also
to ``pure'' $\mathcal{N}=5$ supergravity, which is based on $M_{1,2}\left(
\mathbb{O}\right) $, the Jordan triple system (not upliftable to $d=5$)
generated by $2\times 1$ matrices over $\mathbb{O}$ \cite{GST}.

In the complex field $G_{4}=SU\left( 6\right) $, and $\mathbf{Sympl}\left(
SU\left( 6\right) \right) =\mathbf{20}$, the real self-dual rank-$3$
antisymmetric irreducible representation.

\begin{itemize}
\item  Let us first solve the problem for $p=1$. $\mathcal{G}_{1}$ is a real
form of $SU(3)\times SU(3)$, the relevant (maximal symmetric) embedding is
\begin{equation}
SU(6)\supset SU(3)\times SU(3)\times U(1)=\mathcal{G}_{1}\times U(1),
\label{emb-2}
\end{equation}
and the $\mathbf{20}$ accordingly branches as
\begin{equation}
\mathbf{20}\rightarrow (\mathbf{1},\mathbf{1})_{-3}+(\mathbf{3},\overline{%
\mathbf{3}})_{-1}+(\overline{\mathbf{3}},\mathbf{3})_{+1}+(\mathbf{1},%
\mathbf{1})_{+3}\,,  \label{branchsu6}
\end{equation}
corresponding to the charge decomposition
\begin{equation}
\mathbf{Q}_{1}\rightarrow (p^{0},\mathbf{p_{(\mathbf{3},\overline{\mathbf{3}}%
)}},q_{0},\mathbf{q_{(\overline{\mathbf{3}},\mathbf{3})}})\,.
\end{equation}
The analysis here is analogous to the cases treated above. The branching (%
\ref{branchsu6}) contains two $\left[ SU(3)\times SU(3)\right] $-singlets,
but, by virtue of (\ref{emb-2}), one of the singlets can be freely acted on
by the $U(1)$. By acting with $G_{4}/\mathcal{G}_{1}=SU\left( 6\right) /%
\left[ SU(3)\times SU(3)\right] $, $\mathbf{Q}_{1}$ can be reduced to
\begin{equation}
\mathbf{Q}_{1}\overset{SU\left( 6\right) /\left[ SU(3)\times SU(3)\right] }{%
\longrightarrow }(I^{(1)},\mathbf{0}_{(\mathbf{3},\overline{\mathbf{3}}%
)},\pm I^{(1)},\mathbf{0}_{(\overline{\mathbf{3}},\mathbf{3})})\,,
\end{equation}
so that $I_{1}=1$, which corresponds to $\mathcal{I}_{4}\left( \mathcal{Q}%
_{1}^{4}\right) $ only. Indeed, formula (\ref{conta}) yields
\begin{equation}
I_{1}+\text{dim}_{\mathbb{R}}(SU\left( 6\right) )-\text{dim}_{\mathbb{R}%
}(SU(3)\times SU(3))=1+35-16=20\,.
\end{equation}

\item  Let us consider now the $2$-centered case ($p=2$). Having solved the
problem for $p=1$, we further decompose $\mathbf{Q}_{2}$ with respect to $%
\mathcal{G}_{1}=SU(3)\times SU(3)$:
\begin{equation}
\mathbf{Q}_{2}=(I^{(2)},\mathbf{m_{(\mathbf{3},\overline{\mathbf{3}})}}%
,I^{(3)},\mathbf{e_{(\overline{\mathbf{3}},\mathbf{3})}}),
\end{equation}
and find the corresponding little group. The little group of the $(\mathbf{3}%
,\overline{\mathbf{3}})$ of $SU(3)\times SU(3)$ is the \textit{diagonal} $%
SU(3)$,which is maximal in $SU(3)\times SU(3)$ (see \textit{e.g.} \cite
{Slansky}), under which such a representation branches as follows:
\begin{equation}
(\mathbf{3},\overline{\mathbf{3}})\rightarrow \mathbf{1}+\mathbf{8}\,,
\end{equation}
yielding the charge decompositions
\begin{equation*}
\mathbf{m}_{(\mathbf{3},\overline{\mathbf{3}})}\rightarrow (I^{(4)},\mathbf{m%
}_{8});~~\mathbf{e_{(\overline{\mathbf{3}},\mathbf{3})}}\rightarrow (I^{(5)},%
\mathbf{e}_{8}).
\end{equation*}
The maximality of the embedding of the diagonal $SU(3)$ in $SU(3)\times
SU(3) $ implies all singlets to correspond to extra $SU\left( 6\right) $%
-invariant polynomials, and $\mathbf{m}_{\mathbf{8}}$ can be set to zero
through the action of $\mathcal{G}_{1}/SU(3)=\left[ SU(3)\times SU(3)\right]
/SU(3)$, thus yielding the result:
\begin{equation}
\mathbf{Q}_{2}\overset{\left[ SU(3)\times SU(3)\right] /SU(3)}{%
\longrightarrow }\left( I^{(2)},I^{(4)},\mathbf{0}_{\mathbf{8}%
},I^{(3)},I^{(5)},\mathbf{e}_{\mathbf{8}}\right) .
\end{equation}

\item  The $\mathbf{8}$ (adjoint) of $SU(3)$ has little group $[U(1)]^{2}$,
which does not commute with any $U(1)$ in $SU(3)$. The $\mathbf{8}$
correspondingly branches as
\begin{equation}
\mathbf{8}\rightarrow \mathbf{1}_{0,0}+\mathbf{1}_{0,0}+\mathbf{1}_{0,2}+%
\mathbf{1}_{0,-2}+\mathbf{1}_{3,1}+\mathbf{1}_{3,-1}+\mathbf{1}_{-3,1}+%
\mathbf{1}_{-3,-1}\,,  \label{cern-1}
\end{equation}
and thus
\begin{equation}
\mathbf{e}_{\mathbf{8}}\longrightarrow (I^{(6)},I^{(7)},\mathrm{e}_{0,2},%
\mathrm{e}_{0,-2},\mathrm{e}_{3,1},\mathrm{e}_{3,-1},\mathrm{e}_{-3,1},%
\mathrm{e}_{-3,-1}).
\end{equation}
Therefore, by acting with $SU(3)/\mathcal{G}_{2}=SU(3)/$ $[U(1)]^{2}$, $%
\mathbf{Q}_{2}$ can then be put in the form
\begin{equation}
\mathbf{Q}_{2}\overset{SU(3)/[U(1)]^{2}}{\longrightarrow }(I^{(2)},I^{(4)},%
\mathbf{0}_{\mathbf{8}},I^{(3)},I^{(5)},I^{(6)},I^{(7)},\mathbf{0}_{6}),
\end{equation}
where $\mathbf{0}_{6}$ collectively denotes the six charges pertaining to
the $[U(1)]^{2}$-charged representations $\mathbf{1}_{0,2}$, $\mathbf{1}%
_{0,-2}$, $\mathbf{1}_{3,1}$, $\mathbf{1}_{3,-1}$, $\mathbf{1}_{-3,1}$, $%
\mathbf{1}_{-3,-1}$ in the right-hand side of (\ref{cern-1}).
\end{itemize}

In conclusion, we found that the little group of a $2$-centered black-hole
solution is $\mathcal{G}_{2}=[U(1)]^{2}$, and the corresponding $2$-centered
charge orbit reads (in complexified form)
\begin{equation}
\mathcal{O}_{p=2}=\frac{G_{4}}{\mathcal{G}_{2}}=\frac{SU\left( 6\right) }{%
[U(1)]^{2}}.
\end{equation}
The $SU\left( 6\right) $-invariant polynomials for a $2$-centered
configuration are seven: $I_{2}=7$; indeed, the general formula (\ref{conta}%
) gives:
\begin{equation}
I_{2}+\text{dim}_{\mathbb{R}}(SU\left( 6\right) )-\text{dim}_{\mathbb{R}%
}([U(1)]^{2})=7+35-2=40=2\cdot 20.
\end{equation}

\subsection{\label{J3-R}$J_{3}^{\mathbb{R}}$ ($\mathcal{N}=2$)}

Finally, we consider the model based on the Euclidean degree-$3$ Jordan
algebra $J_{3}^{\mathbb{R}}$ on $\mathbb{R}$.

In the complex field $G_{4}=USp\left( 6\right) $, and $\mathbf{Sympl}\left(
USp\left( 6\right) \right) =\mathbf{14}^{\prime }$, the real self-dual rank-$%
3$ antisymmetric irreducible representation of $USp\left( 6\right) $ (not to
be confused with the rank-$2$ antisymmetric irreducible representation $%
\mathbf{14}$ considered in Section \ref{J3-H}).

\begin{itemize}
\item  Let us first solve the problem for $p=1$. $\mathcal{G}_{1}$ is a real
form of $SU(3)$, the relevant (maximal symmetric) embedding is
\begin{equation}
USp(6)\supset SU(3)\times U(1)=\mathcal{G}_{1}\times U(1),  \label{emb-3}
\end{equation}
and the $\mathbf{14}^{\prime }$ accordingly branches as
\begin{equation}
\mathbf{14}^{\prime }\rightarrow \mathbf{1}_{-3}+\mathbf{6}_{-1}+\overline{%
\mathbf{6}}_{+1}+\mathbf{1}_{+3},\,  \label{branchusp6}
\end{equation}
corresponding to the charge decomposition
\begin{equation}
\mathbf{Q}_{1}\rightarrow (p^{0},\mathbf{p}_{\mathbf{6}},q_{0},\mathbf{q}_{%
\mathbf{6}})\,.
\end{equation}
Once again, the analysis here is analogous to the cases treated above. The
branching (\ref{branchusp6}) contains two $SU(3)$-singlets, but, by virtue
of (\ref{emb-3}), one of the singlets can be freely acted on by the $U(1)$.
By acting with $G_{4}/\mathcal{G}_{1}=USp\left( 6\right) /SU(3)$, $\mathbf{Q}%
_{1}$ can be reduced to
\begin{equation}
\mathbf{Q}_{1}\overset{USp\left( 6\right) /SU(3)}{\longrightarrow }(I^{(1)},%
\mathbf{0}_{\mathbf{6}},\pm I^{(1)},\mathbf{0}_{\mathbf{6}})\,,
\end{equation}
so that $I_{1}=1$, which corresponds to $\mathcal{I}_{4}\left( \mathcal{Q}%
_{1}^{4}\right) $ only. Indeed, formula (\ref{conta}) yields
\begin{equation}
I_{1}+\text{dim}_{\mathbb{R}}(USp\left( 6\right) )-\text{dim}_{\mathbb{R}%
}(SU(3))=1+21-8=14\,.
\end{equation}

\item  Let us consider now the $2$-centered case ($p=2$). Having solved the
problem for $p=1$, we further decompose $\mathbf{Q}_{2}$ with respect to $%
\mathcal{G}_{1}=SU(3)$:
\begin{equation}
\mathbf{Q}_{2}=(I^{(2)},\mathbf{m}_{\mathbf{6}},I^{(3)},\mathbf{e}_{\mathbf{6%
}}),
\end{equation}
and find the corresponding little group. The little group of the $\mathbf{6}$
of $SU(3)$ is $SO(3)$,which is maximal in $SU(3)$, under which such a
representation branches as follows:
\begin{equation}
\mathbf{6}\rightarrow \mathbf{1}+\mathbf{5},
\end{equation}
yielding the charge decompositions
\begin{equation}
\mathbf{m}_{\mathbf{6}}\rightarrow (I^{(4)},\mathbf{m}_{\mathbf{5}});~~%
\mathbf{e}_{\mathbf{6}}\rightarrow (I^{(5)},\mathbf{e}_{\mathbf{5}}).
\end{equation}
The maximality of $SO(3)$ in $SU(3)$ implies all singlets to corresponds to
extra $USp\left( 6\right) $-invariant polynomials, and $\mathbf{m}_{\mathbf{6%
}}$ can be set to zero through the action of $\mathcal{G}%
_{1}/SO(3)=SU(3)/SO(3)$, thus yielding the result:
\begin{equation}
\mathbf{Q}_{2}\overset{SU(3)/SO(3)}{\longrightarrow }(I^{(2)},I^{(4)},%
\mathbf{0}_{\mathbf{5}},I^{(3)},I^{(5)},\mathbf{e}_{\mathbf{5}}).
\end{equation}

\item  Note, however, that the little group of the $\mathbf{5}$ (rank-$2$
symmetric traceless) irreducible representation of $SO(3)$ is the identity,
so that $\mathcal{G}_{2}=\mathbb{I}$. The $\mathbf{5}$ then trivially
branches into five singlets, three of which can be rotated to zero through
the action of $SO(3)/\mathcal{G}_{2}=SO(3)$:
\begin{equation}
\mathbf{Q}_{2}\overset{SO(3)}{\longrightarrow }(I^{(2)},I^{(4)},\mathbf{0}_{%
\mathbf{5}},I^{(3)},I^{(5)},I^{(6)},I^{(7)},\mathbf{0}_{3}),
\end{equation}
where $\mathbf{0}_{3}$ collectively denotes such three singlets set to zero.
\end{itemize}

In conclusion, we found that the little group of a $2$-centered black-hole
solution is the identity itself: $\mathcal{G}_{2}=\mathbb{I}$, and the
corresponding $2$-centered charge orbit reads (in compact form)
\begin{equation}
\mathcal{O}_{p=2}=\frac{G_{4}}{\mathcal{G}_{2}}=USp\left( 6\right) .
\end{equation}
The $USp\left( 6\right) $-invariant polynomials for a $2$-centered
configuration are seven: $I_{2}=7$; indeed, the general formula (\ref{conta}%
) yields:
\begin{equation}
I_{2}+\text{dim}_{\mathbb{R}}(USp(6))-\text{dim}_{\mathbb{R}}(\mathbb{I}%
)=7+21-0=28=2\cdot 14.
\end{equation}

\section{\label{Formal-Analysis}Invariant Structures and the role of the
Horizontal Symmetry $SL_{h}\left( 2,\mathbb{R}\right) $}

We now propose a candidate for a complete basis of $G_{4}$-invariant
polynomials for the $p=2$ case, highlighting the role of the horizontal
symmetry group \cite{FMOSY-1} in the classification of multi-center
invariant structures.

Our treatment applies at least to the \textit{irreducible} cubic geometries
of \textit{symmetric} scalar manifolds of $d=4$ supergravity theories \cite
{dWVVP} (which, with the exception of the rank-1 $t^{3}$ model\footnote{%
As mentioned above, the irreducible rank-$1$ cubic case (the so-called $%
\mathcal{N}=2$, $d=4$ $t^{3}$ model, associated to the trivial degree-$1$
Jordan algebra $\mathbb{R}$) has been treated in \cite{FMOSY-1}.}, are the
ones considered in the counting analysis of Section \ref{Counting-Analysis}):

\begin{enumerate}
\item  $\mathcal{N}=2$ magical Maxwell-Einstein supergravities ($J_{3}^{%
\mathbb{A}}$, $\mathbb{A}=\mathbb{O},\mathbb{H},\mathbb{C},\mathbb{R}$),
with the case $J_{3}^{\mathbb{H}}$ encompassing also $\mathcal{N}=6$
``pure'' supergravity \cite{GST,ADF,FGimK,ADFGT,RS};

\item  $\mathcal{N}=5$ ``pure'' supergravity ($M_{1,2}\left( \mathbb{O}%
\right) $);

\item  $\mathcal{N}=8$ ``pure'' supergravity ($J_{3}^{\mathbb{O}_{s}}$).
\end{enumerate}

The simplest invariant structures of a simple Lie group $G$ (such as the $U$%
-duality group $G_{4}$ of an \textit{irreducible} symmetric model) are the
Killing-Cartan metric $g_{\alpha \beta }$, the structure constants $%
f_{\alpha \beta \gamma }$ and the symplectic metric $\mathbb{C}_{MN}$ (the
Greek indices are in the adjoint representation of $G_4$, $\mathbf{Adj}%
\left( G_{4}\right) $, while the capital indices are in $\mathbf{Sympl}%
\left( G_{4}\right) $). It is well known that the entries of the generators
in $\mathbf{Sympl}\left( G_{4}\right) $
\begin{equation}
t_{\alpha \mid MN}\equiv t_{\alpha \mid M}{}^{P}\mathbb{C}_{PN}\,=t_{\alpha
\mid \left( MN\right) }  \label{gen}
\end{equation}
are invariant structures, symmetric in the symplectic indices (for the
notation, see \cite{deWit:2002vt}).

In particular, one can construct the so-called $K$-tensor\footnote{%
With respect to the treatment given in \cite{Exc-Reds}, we fix the overall
normalization constant of the $K$-tensor to the value $\xi =-\frac{1}{3\tau }%
=-\frac{f\left( f+1\right) }{6d}$, as needed for consistency reasons.} \cite
{Exc-Reds}
\begin{equation}
\mathbb{K}_{MNPQ}\equiv -\frac{1}{3\tau }t_{(MN}^{\alpha }t_{\alpha |PQ)}=-%
\frac{1}{3\tau }\left( t_{MN}^{\alpha }t_{\alpha |PQ}-\tau \,\mathbb{C}_{M(P}%
\mathbb{C}_{Q)N}\right) \,=\mathbb{K}_{\left( MNPQ\right) },  \label{K-t}
\end{equation}
where $\tau $ is a $G_{4}$-dependent constant defined as
\begin{equation}
\tau \equiv \frac{2d}{f(f+1)}\,,  \label{tau}
\end{equation}
with $d\equiv $dim$_{\mathbb{R}}\mathbf{Adj}\left( G_{4}\right) $ and $%
f\equiv $dim$_{\mathbb{R}}\left( \mathbf{Sympl}\left( G_{4}\right) \right) $%
.\medskip\ From its definition (\ref{K-t}), the $K$-tensor is a completely
symmetric rank-$4$ $G_{4}$-invariant tensor of $\mathbf{Sympl}\left(
G_{4}\right) $.

In the presence of a single-centered black-hole background ($p=1$),
associated to a dyonic black-hole charge vector $\mathcal{Q}^{M}$ in $%
\mathbf{Sympl}\left( G_{4}\right) $, the unique independent $G_{4}$%
-invariant polynomial reads \cite{Exc-Reds}
\begin{equation}
\mathcal{I}_{4}\left( \mathcal{Q}^{4}\right) \equiv \mathbb{K}_{MNPQ}%
\mathcal{Q}^{M}\mathcal{Q}^{N}\mathcal{Q}^{P}\mathcal{Q}^{Q}=-\frac{1}{3\tau
}t_{MN}^{\alpha }t_{\alpha |PQ}\mathcal{Q}^{M}\mathcal{Q}^{N}\mathcal{Q}^{P}%
\mathcal{Q}^{Q}.  \label{I4-1-center-def}
\end{equation}

On the other hand, in the presence of a multi-centered black-hole solution ($%
p\geqslant 2$), the horizontal symmetry $SL_{h}\left( p,\mathbb{R}\right) $
\cite{FMOSY-1} plays a crucial role in organizing the various $G_{4}$%
-covariant and $G_{4}$-invariant structures.

In the following treatment we will consider the $2$-centered case ($p=2$),
the index $a=1,2$ spanning the fundamental representation (spin $s=1/2$) $%
\mathbf{2} $ of the horizontal symmetry $SL_{h}\left( 2,\mathbb{R}\right) $.

By using the symplectic representation (\ref{gen}) of the generators of $%
G_{4}$, one can introduce the tensor (homogeneous quadratic in charges)
\begin{equation}
T_{\alpha \mid ab}\equiv t_{\alpha |MN}\mathcal{Q}_{a}^{M}\mathcal{Q}%
_{b}^{N}\,=T_{\alpha \mid \left( ab\right) }=\left(
\begin{array}{cc}
T_{\alpha \mid 11} & T_{\alpha \mid 12} \\
T_{\alpha \mid 12} & T_{\alpha \mid 22}
\end{array}
\right) ,  \label{def-T}
\end{equation}
lying in $\left( \mathbf{3},\mathbf{Adj}\left( G_{4}\right) \right) $ of $%
SL_{h}\left( 2,\mathbb{R}\right) \times G_{4}$, where $\mathbf{3}$ is the
rank-$2$ symmetric (spin $s=1$) representation of $SL_{h}\left( 2,\mathbb{R}%
\right) $. In \textit{irreducible} models, $T_{\alpha \mid ab}$ is the
analogue of the so-called $\mathbb{T}$-tensor, introduced in \cite{FMOSY-1}\
for \textit{reducible} theories. Under the centers' exchange $%
1\leftrightarrow 2$, $T_{\alpha \mid 11}\leftrightarrow T_{\alpha \mid 22}$,
while $T_{\alpha \mid 12}$ is invariant.

Interestingly, one can prove that the quantity
\begin{equation}
\mathbf{N}\equiv g^{\alpha \beta }\left( T_{\alpha \mid 11}T_{\beta \mid
22}-T_{\alpha \mid 12}T_{\beta \mid 12}\right)  \label{def-N}
\end{equation}
is \textit{not} independent from lower order invariants. Indeed, \textit{at
least} in the aforementioned irreducible cases, it holds that
\begin{equation}
t_{M[N}^{\alpha }t_{\alpha \mid P]Q}=\frac{\tau }{2}\left[ \mathbb{C}_{M(P}%
\mathbb{C}_{Q)N}-\mathbb{C}_{M(N}\mathbb{C}_{Q)P}\right] .  \label{2}
\end{equation}
Thus, from (\ref{def-T}) and (\ref{def-N}), it follows that
\begin{equation}
\mathbf{N}=2t_{M[N}^{\alpha }t_{\alpha \mid P]Q}\mathcal{Q}_{1}^{M}\mathcal{Q%
}_{1}^{N}\mathcal{Q}_{2}^{P}\mathcal{Q}_{2}^{Q}=-\frac{1}{3}\left[ \mathbb{C}%
_{M(P}\mathbb{C}_{Q)N}-\mathbb{C}_{M(N}\mathbb{C}_{Q)P}\right] \mathcal{Q}%
_{1}^{M}\mathcal{Q}_{1}^{N}\mathcal{Q}_{2}^{P}\mathcal{Q}_{2}^{Q}=\frac{1}{2}%
\mathcal{W}^{2},  \label{3}
\end{equation}
where
\begin{equation}
\mathcal{W}\equiv \left\langle \mathcal{Q}_{1},\mathcal{Q}_{2}\right\rangle
\equiv \frac{1}{2}\mathbb{C}_{MN}\epsilon ^{ab}\mathcal{Q}_{a}^{M}\mathcal{Q}%
_{b}^{N}  \label{4}
\end{equation}
is the \textit{symplectic product} of the charge vectors $\mathcal{Q}_{1}$
and $\mathcal{Q}_{2}$, which is a singlet $\left( \mathbf{1},\mathbf{1}%
\right) $ of $SL_{h}\left( 2,\mathbb{R}\right) \times G_{4}$ (manifestly
antisymmetric under $1\leftrightarrow 2$).

An important difference between the \textit{reducible} models (studied in
\cite{FMOSY-1}) and the \textit{irreducible} treated in the present
investigation is that, while the former generally have a non-vanishing
horizontal invariant polynomial $\mathcal{X}$, the latter have it vanishing
identically. Indeed, the analogue of $\mathcal{X}$ (defined by Eq. (4.13) of
\cite{FMOSY-1}) for irreducible models can be defined as
\begin{equation}
\mathcal{X}_{irred}\equiv \mathbf{N}-\frac{1}{2}\mathcal{W}^{2}=0,
\label{Xcall-irred}
\end{equation}
where result (\ref{3}) was used in the last step. The $t^{3}$ model
mentioned in the Introduction is a non-generic irreducible model (studied in
Sec. 7 of \cite{FMOSY-1}); in this case, the vanishing of $\mathcal{X}$ is
given by Eq. (7.16) of \cite{FMOSY-1}.

By using the $K$-tensor (\ref{K-t}), one can also define the tensor
(homogeneous cubic in charges)
\begin{equation}
\mathcal{Q}_{M\mid abc}\equiv \mathbb{K}_{MNPQ}\mathcal{Q}_{a}^{N}\mathcal{Q}%
_{b}^{P}\mathcal{Q}_{c}^{Q}=\mathcal{Q}_{M\mid \left( abc\right) },
\label{Q-def}
\end{equation}
lying in $\left( \mathbf{4},\mathbf{Sympl}\left( G_{4}\right) \right) $ of $%
SL_{h}\left( 2,\mathbb{R}\right) \times G_{4}$, where $\mathbf{4}$ is the
rank-$3$ symmetric representation (spin $s=3/2$) of $SL_{h}\left( 2,\mathbb{R%
}\right) $. Under $1\leftrightarrow 2$, it holds that $\mathcal{Q}_{M\mid
111}\leftrightarrow \mathcal{Q}_{M\mid 222}$ and $\mathcal{Q}_{M\mid
112}\leftrightarrow \mathcal{Q}_{M\mid 122}$.

By further contracting with a $2$-centered charge vector, one can introduce
the tensor (homogeneous quartic in charges)
\begin{equation}
\mathbf{I}_{abcd}\equiv \mathbb{K}_{MNPQ}\mathcal{Q}_{a}^{M}\mathcal{Q}%
_{b}^{N}\mathcal{Q}_{c}^{P}\mathcal{Q}_{d}^{Q}=\mathbf{I}_{\left(
abcd\right) },  \label{I-def}
\end{equation}
lying in $\left( \mathbf{5},\mathbf{1}\right) $ of $SL_{h}\left( 2,\mathbb{R}%
\right) \times G_{4}$, where $\mathbf{5}$ is the rank-$4$ symmetric
representation (spin $s=2$) of $SL_{h}\left( 2,\mathbb{R}\right) $. Under $%
1\leftrightarrow 2$, $\mathbf{I}_{1111}\leftrightarrow \mathbf{I}_{2222}$, $%
\mathbf{I}_{1112}\leftrightarrow \mathbf{I}_{1222}$, while $\mathbf{I}%
_{1122} $ is invariant.

Trivially, $\widetilde{\mathcal{Q}}_{abc}\equiv \mathcal{Q}_{M\mid abc}$ and
$\mathbf{I}_{\left( abcd\right) }$ are related by\footnote{%
We remark that relation (\ref{rel-2}) characterizes $\widetilde{\mathcal{Q}}%
_{abc}$ as the $2$-center generalisation of the so-called \textit{%
Freudenthal dual} of the dyonic charge vector $\mathcal{Q}^{M}$, introduced
(with a different normalisation) in \cite{Duff-Freud}. Thus, $\widetilde{%
\mathcal{Q}}_{abc}$ can be regarded as the (polynomial) $2$\textit{-center
Freudenthal dual} of the dyonic charge vector $\mathcal{Q}_{d}$.
\par
Furthermore, Eqs. (\ref{4}), (\ref{rel-1}) and (\ref{I6}) yield that, under
the formal interchange $\mathcal{Q}_{a}^{M}\leftrightarrow \mathbb{C}^{MN}%
\mathcal{Q}_{N\mid abc}$, $\mathbf{I}_{abcd}$ is invariant and $\mathcal{W}%
\leftrightarrow \mathbf{I}_{6}$.}
\begin{eqnarray}
\mathbf{I}_{abcd} &=&\mathcal{Q}_{M\mid abc}\mathcal{Q}_{d}^{M}=\mathbb{C}%
^{MN}\mathcal{Q}_{M\mid abc}\mathcal{Q}_{N\mid d}=\left\langle \widetilde{%
\mathcal{Q}}_{abc},\mathcal{Q}_{d}\right\rangle ;  \label{rel-1} \\
\mathcal{Q}_{M\mid abc} &=&\frac{1}{4}\frac{\partial \mathbf{I}_{abcd}}{%
\partial \mathcal{Q}_{d}^{M}}.  \label{rel-2}
\end{eqnarray}
Note that only the completely symmetric part $\mathcal{Q}_{M\mid (abc}%
\mathcal{Q}_{d)}^{M}$ survives the contraction in (\ref{rel-1}), because $%
\mathcal{Q}_{M\mid abc}\mathcal{Q}_{d}^{M}\epsilon ^{cd}=0$ from the
symmetry of the $K$-tensor (\ref{K-t}) and the definition (\ref{Q-def}) of $%
\mathcal{Q}_{M\mid abc}$ itself.

In order to generate $G_{4}$-invariant polynomials, one can:

\begin{enumerate}
\item  multiply and contract on $\mathbf{Adj}\left( G_{4}\right) $ the three
components of the quadratic tensor $T_{\alpha \mid ab}$ defined by (\ref
{def-T}), \textit{or}

\item  contract all four components of $\mathcal{Q}_{M\mid abc}$ defined by (%
\ref{Q-def}) with three $2$-center charge vectors, in all possible ways,
\textit{or}

\item  contract all five components of $\mathbf{I}_{abcd}$ defined by (\ref
{I-def}) with four $2$-center charge vectors, in all possible ways.
\end{enumerate}

By virtue of the various relations considered above, these three approaches
give equivalent results, which we now specify for the sake of clarity:
\begin{eqnarray}
\mathbf{I}_{+2}\left( \mathcal{Q}_{1}^{4}\right) &\equiv &\mathcal{I}%
_{4}\left( \mathcal{Q}_{1}^{4}\right) \equiv \mathbf{I}_{1111}=\left\langle
\widetilde{\mathcal{Q}}_{111},\mathcal{Q}_{1}\right\rangle =\mathbb{K}_{MNPQ}%
\mathcal{Q}_{1}^{M}\mathcal{Q}_{1}^{N}\mathcal{Q}_{1}^{P}\mathcal{Q}%
_{1}^{Q}=-\frac{1}{3\tau }T_{11}^{\alpha }T_{\alpha \mid 11};  \notag \\
&&  \label{I+2} \\
\mathbf{I}_{+1}\left( \mathcal{Q}_{1}^{3}\mathcal{Q}_{2}\right) &\equiv &%
\mathbf{I}_{1112}=\left\langle \widetilde{\mathcal{Q}}_{111},\mathcal{Q}%
_{2}\right\rangle =\left\langle \widetilde{\mathcal{Q}}_{112},\mathcal{Q}%
_{1}\right\rangle =\mathbb{K}_{MNPQ}\mathcal{Q}_{1}^{M}\mathcal{Q}_{1}^{N}%
\mathcal{Q}_{1}^{P}\mathcal{Q}_{2}^{Q}=-\frac{1}{3\tau }T_{11}^{\alpha
}T_{12\mid \alpha };  \notag \\
&&  \label{I+1} \\
\mathbf{I}_{0}\left( \mathcal{Q}_{1}^{2}\mathcal{Q}_{2}^{2}\right) &\equiv &%
\mathbf{I}_{1122}=\left\langle \widetilde{\mathcal{Q}}_{112},\mathcal{Q}%
_{2}\right\rangle =\left\langle \widetilde{\mathcal{Q}}_{122},\mathcal{Q}%
_{1}\right\rangle =\mathbb{K}_{MNPQ}\mathcal{Q}_{1}^{M}\mathcal{Q}_{1}^{N}%
\mathcal{Q}_{2}^{P}\mathcal{Q}_{2}^{Q}  \notag \\
&=&-\frac{1}{9\tau }\left( T_{11}^{\alpha }T_{22\mid \alpha
}+2T_{12}^{\alpha }T_{12\mid \alpha }\right) =-\frac{1}{3\tau }\left(
T_{11}^{\alpha }T_{22\mid \alpha }+\tau \mathcal{W}^{2}\right) ;  \label{I0}
\\
&&  \notag \\
\mathbf{I}_{-1}\left( \mathcal{Q}_{1}\mathcal{Q}_{2}^{3}\right) &\equiv &%
\mathbf{I}_{1222}=\left\langle \widetilde{\mathcal{Q}}_{122},\mathcal{Q}%
_{2}\right\rangle =\left\langle \widetilde{\mathcal{Q}}_{222},\mathcal{Q}%
_{1}\right\rangle =\mathbb{K}_{MNPQ}\mathcal{Q}_{1}^{M}\mathcal{Q}_{2}^{N}%
\mathcal{Q}_{2}^{P}\mathcal{Q}_{2}^{Q}=-\frac{1}{3\tau }T_{22}^{\alpha
}T_{12\mid \alpha };  \notag \\
&&  \label{I-1} \\
\mathbf{I}_{-2}\left( \mathcal{Q}_{2}^{4}\right) &\equiv &\mathcal{I}%
_{4}\left( \mathcal{Q}_{2}^{4}\right) \equiv \mathbf{I}_{2222}=\left\langle
\widetilde{\mathcal{Q}}_{222},\mathcal{Q}_{2}\right\rangle =\mathbb{K}_{MNPQ}%
\mathcal{Q}_{2}^{M}\mathcal{Q}_{2}^{N}\mathcal{Q}_{2}^{P}\mathcal{Q}%
_{2}^{Q}=-\frac{1}{3\tau }T_{22}^{\alpha }T_{22\mid \alpha }.  \label{I-2}
\end{eqnarray}
The subscripts in the $G_{4}$-invariant polynomials $\mathbf{I}_{+2}$, $%
\mathbf{I}_{+1}$, $\mathbf{I}_{0}$, $\mathbf{I}_{-1}$ and $\mathbf{I}_{-2}$
defined by (\ref{I+2})-(\ref{I-2}) denote the polarization with respect to
the horizontal symmetry $SL_{h}\left( 2,\mathbb{R}\right) $, inherited from
the components of $\mathbf{I}_{abcd}$ (\ref{I-def}); indeed, the five $G_{4}$%
-invariant polynomials (\ref{I+2})-(\ref{I-2}) sit in the rank-$4$ symmetric
representation (spin $s=2$) $\mathbf{5}$ of $SL_{h}\left( 2,\mathbb{R}%
\right) $ itself \cite{FMOSY-1}.

In order to proceed further, it is worth mentioning the decomposition \cite
{Exc-Reds}
\begin{equation}
t_{\alpha |M}^{\phantom{\alpha M}N}t_{\beta |NQ}=-t_{\alpha |MP}t_{\beta |NQ}%
\mathbb{C}^{PN}=\frac{1}{2n}g_{\alpha \beta }\mathbb{C}_{MQ}+\frac{1}{2}%
f_{\alpha \beta }{}^{\gamma }\,t_{\gamma |MQ}+S_{(\alpha \beta )[MQ]}\,,
\label{ttDec}
\end{equation}
where
\begin{equation}
S_{\alpha \beta \mid MN}=S_{\left( \alpha \beta \right) \mid \left[ MN\right]
}
\end{equation}
denotes an invariant primitive tensor of $G_{4}$. From (\ref{ttDec}), the
following identity for the $K$-tensor can be derived \cite{Exc-Reds} (recall
Footnote 6):
\begin{eqnarray}
&&\mathbb{K}_{MNPQ}\mathbb{K}_{RSTU}\mathbb{C}^{QR}=-\frac{\left( f+1\right)
}{6d}\mathbb{K}_{\left( MN\right| (ST}\mathbb{C}_{U)\left| P\right) }+\frac{%
(f+1)}{18d}\mathbb{C}_{\left( M\right| \left( S\right| }\mathbb{C}_{\left|
N\right| \left| T\right| }\mathbb{C}_{\left| P\right) \left| U\right) }
\notag \\
&&+\frac{f^{2}\left( f+1\right) ^{2}}{72d^{2}}f_{\alpha \beta \gamma
}t_{~(MN}^{\alpha }t_{~P)(S}^{\beta }t_{~TU)}^{\gamma }-\frac{f^{2}\left(
f+1\right) ^{2}}{36d^{2}}t_{~(MN}^{\alpha }S_{\alpha \beta \mid
P)(S}t_{~TU)}^{\beta }\,,  \label{masteridentity}
\end{eqnarray}
where
\begin{equation}
S_{\alpha \beta \mid 12}\equiv S_{\alpha \beta \mid MN}\mathcal{Q}_{1}^{M}%
\mathcal{Q}_{2}^{N}=S_{\alpha \beta \mid MN}\mathcal{Q}_{1}^{[M}\mathcal{Q}%
_{2}^{N]}=-S_{\alpha \beta \mid 21}.
\end{equation}

A $G_{4}$-invariant polynomial homogeneous sextic in charges can then be
defined as follows:
\begin{eqnarray}
\mathbf{I}_{6}\left( \mathcal{Q}_{1}^{3}\mathcal{Q}_{2}^{3}\right) &\equiv &%
\frac{1}{8}\left\langle \widetilde{\mathcal{Q}}_{abc},\widetilde{\mathcal{Q}}%
_{def}\right\rangle \epsilon ^{ad}\epsilon ^{be}\epsilon ^{cf}=\frac{1}{8}%
\mathbb{C}^{MN}\mathcal{Q}_{M\mid abc}\mathcal{Q}_{N\mid def}\epsilon
^{ad}\epsilon ^{be}\epsilon ^{cf}  \notag \\
&=&\frac{1}{4}\left\langle \widetilde{\mathcal{Q}}_{111},\widetilde{\mathcal{%
Q}}_{222}\right\rangle +\frac{3}{4}\left\langle \widetilde{\mathcal{Q}}%
_{122},\widetilde{\mathcal{Q}}_{112}\right\rangle  \notag \\
&=&\frac{1}{4}\mathbb{K}_{MNPQ}\mathbb{K}_{RSTU}\mathbb{C}^{QR}\left(
\mathcal{Q}_{1}^{M}\mathcal{Q}_{1}^{N}\mathcal{Q}_{1}^{P}\mathcal{Q}_{2}^{S}%
\mathcal{Q}_{2}^{T}\mathcal{Q}_{2}^{U}+3\mathcal{Q}_{1}^{M}\mathcal{Q}%
_{2}^{N}\mathcal{Q}_{2}^{P}\mathcal{Q}_{1}^{S}\mathcal{Q}_{1}^{T}\mathcal{Q}%
_{2}^{U}\right)  \notag \\
&=&\frac{\left( f+1\right) }{36d}\mathcal{W}^{3}+\frac{f^{2}\left(
f+1\right) ^{2}}{144d^{2}}f_{\alpha \beta \gamma }T_{11}^{\alpha
}T_{12}^{\beta }T_{22}^{\gamma }+\frac{f^{2}\left( f+1\right) ^{2}}{108d^{2}}%
\left( T_{12}^{\alpha }T_{12}^{\beta }-T_{11}^{\alpha }T_{22}^{\beta
}\right) S_{\alpha \beta \mid 12}.  \notag \\
&&  \label{I6}
\end{eqnarray}
Note that $\mathbf{I}_{6}$ is manifestly antisymmetric under $%
1\leftrightarrow 2$. The first line of (\ref{I6}) is manifestly $\left[
SL_{h}\left( 2,\mathbb{R}\right) \times G_{4}\right] $-invariant, the second
and third lines provide explicit expressions, and in the fourth line the
``master'' identity (\ref{masteridentity}) was exploited.

If the symplectic product $\mathcal{W}\neq 0$ (defined in (\ref{4})), the
two charge vectors $\mathcal{Q}_{1}^{M}$ and $\mathcal{Q}_{2}^{M}$ are
\textit{mutually non-local}. The concept of \textit{mutual non-locality} is
very important in the treatment of marginal stability in multi-center black
holes (see \textit{e.g.} \cite{D-1,D-2,BD,DM-1,G-1,GLS-2,CS}).

The above treatment suggests that a candidate for a complete basis of $G_{4}$%
-invariant polynomials in the irreducible cases under consideration is given
by the seven polynomials:
\begin{equation}
\left( \mathcal{W},~\mathbf{I}_{+2},~\mathbf{I}_{+1},~\mathbf{I}_{0},~%
\mathbf{I}_{-1},~\mathbf{I}_{-2},~\mathbf{I}_{6}\right) ,  \label{seven?}
\end{equation}
respectively defined by (\ref{4}), (\ref{I+2})-(\ref{I-2}) and (\ref{I6}).
The corresponding candidate for a complete basis of $\left[ SL_{h}\left( 2,%
\mathbb{R}\right) \times G_{4}\right] $-invariant polynomials in the
irreducible cases under consideration is then given by the four polynomials
\begin{equation}
\left( \mathcal{W},~\mathbf{I}_{6},~\text{Tr}\left( \mathbf{I}^{2}\right) ,~%
\text{Tr}\left( \mathbf{I}^{3}\right) \right) ,  \label{four?}
\end{equation}
where \cite{FMOSY-1}
\begin{eqnarray}
\text{Tr}\left( \mathbf{I}^{2}\right) &=&\mathbf{I}_{+2}\mathbf{I}_{-2}+3%
\mathbf{I}_{0}^{2}-4\mathbf{I}_{+1}\mathbf{I}_{-1};  \label{Tr(I^2)} \\
\text{Tr}\left( \mathbf{I}^{3}\right) &=&\mathbf{I}_{0}^{3}+\mathbf{I}_{+2}%
\mathbf{I}_{-1}^{2}+\mathbf{I}_{-2}\mathbf{I}_{+1}^{2}-\mathbf{I}_{+2}%
\mathbf{I}_{-2}\mathbf{I}_{0}-2\mathbf{I}_{+1}\mathbf{I}_{0}\mathbf{I}_{-1}.
\label{Tr(I^3)}
\end{eqnarray}
Indeed, the spin $s=2$ representation $\mathbf{5}$ of $SL_{h}\left( 2,%
\mathbb{R}\right) $, whose components are the $G_{4}$-invariant polynomials $%
\mathbf{I}_{+2}$, $\mathbf{I}_{+1}$, $\mathbf{I}_{0}$, $\mathbf{I}_{-1}$ and
$\mathbf{I}_{-2}$ (defined by (\ref{I+2})-(\ref{I-2})), can be re-arranged
as a $3\times 3$ symmetric traceless matrix $\mathbf{I}$ \cite{FMOSY-1}. (%
\ref{Tr(I^2)}) and (\ref{Tr(I^3)}) (respectively homogeneous of order eight
and twelve in charges) are the only independent $SL_{h}\left( 2,\mathbb{R}%
\right) $-singlets which can be built out of such a $3\times 3$ symmetric
matrix $\mathbf{I}$, due to its tracelessness \cite{FMOSY-1}. Note that Tr$%
\left( \mathbf{I}^{2}\right) $ and Tr$\left( \mathbf{I}^{3}\right) $ are
both invariant under $1\leftrightarrow 2$.\medskip

It is worth pointing out that the analysis of Secs. \ref{Counting-Analysis}
and \ref{Formal-Analysis} can be easily generalised to $p\geqslant 3$
centers. The two-centered representation of spin $s=J/2$ of $SL_{h}\left( 2,%
\mathbb{R}\right) $ is then replaced by the completely symmetric rank-$J$
tensor representation $\mathcal{R}_{J}$ of $SL_{h}\left( p,\mathbb{R}\right)
$ ($J=1,2,3,4$ are the values relevant for the above analysis). On the other
hand, $\mathcal{W}$ and $\mathbf{I}_{6}$ generally sit in the $\left(
\widetilde{\mathcal{R}}_{2},\mathbf{1}\right) $ representation of $%
SL_{h}\left( p,\mathbb{R}\right) \times G_{4}$, where $\widetilde{\mathcal{R}%
}_{2}$ is the rank-$2$ antisymmetric representation of $SL_{h}\left( p,%
\mathbb{R}\right) $ (which, in the case $p=2$, becomes a singlet). However,
due to the tree structure of the split flow in multi-center supergravity
solutions \cite{D-1,D-2,BD,DM-1}, to consider only the case $p=2$ does not
imply any loss in generality (as far as marginal stability issues are
concerned).

\section{\label{Non-Compact}Two-Centered Orbits with Non-Compact Stabiliser:%
\newline
the $\mathcal{N}=8$ BPS and Octonionic $\mathcal{N}=2$ non-BPS Cases}

For $\mathcal{N}=2$ BPS two-centered extremal black holes, the stabiliser of
the supporting charge orbit is always \textit{compact}, so the orbit is
unique (see Table 1 for magical models). This is no longer the case when the
stabiliser is non-compact, as it holds for $\mathcal{N}=2$ two-centered
solutions with two non-BPS centers characterised by $\mathcal{I}_{4}\left(
\mathcal{Q}_{1}^{4}\right) >0$ and $\mathcal{I}_{4}\left( \mathcal{Q}%
_{2}^{4}\right) >0$, and for $\mathcal{N}\geqslant 3$ two-centered solutions
with two $\frac{1}{\mathcal{N}}$-BPS centers. These are interesting cases,
in which a \textit{split attractor flow} through a wall of marginal
stability has been shown to occur \cite{Bossard,MS-FM-1}.

We will consider here the $\frac{1}{8}$- BPS two-centered orbits in the
maximal $\mathcal{N}=8$ theory (based on $J_{3}^{\mathbb{O}_{s}}$) and the
non-BPS two-centered orbits (of the aforementioned type) in the exceptional $%
\mathcal{N}=2$ magic model, based on $J_{3}^{\mathbb{O}}$. These two cases
can be obtained by repeating the analysis of Section \ref{J3-O} and choosing
suitable non-compact real forms of $G_{4}$ and $\mathcal{G}_{2}$.

The $1$-centered charge orbits respectively read \cite{FG1,BFGM1}:
\begin{eqnarray}
\mathcal{N} &=&8,~\frac{1}{8}\text{-BPS}:\mathcal{O}_{p=1}=\frac{E_{7\left(
7\right) }}{E_{6\left( 2\right) }}; \\
\mathcal{N} &=&2,~J_{3}^{\mathbb{O}}~\text{nBPS~}\mathcal{I}_{4}>0:\mathcal{O%
}_{p=1}=\frac{E_{7\left( -25\right) }}{E_{6\left( -14\right) }}.
\end{eqnarray}
In the maximal case, the chain of relevant group branchings reads
\begin{equation}
\mathcal{N}=8,~\frac{1}{8}\text{-BPS}:E_{7\left( 7\right) }\longrightarrow
E_{6\left( 2\right) }\longrightarrow F_{4\left( 4\right) }\longrightarrow
SO\left( 5,4\right) \longrightarrow \left\{
\begin{array}{l}
SO\left( 4,4\right) \\
\mathit{or} \\
SO\left( 5,3\right)
\end{array}
\right. \,,
\end{equation}
such that two $\frac{1}{8}$-BPS, $\mathcal{N}=8$, $2$-centered charge orbits
exist:
\begin{eqnarray}
\mathcal{O}_{\mathcal{N}=8,\frac{1}{8}\text{-BPS},p=2,\mathbf{I}} &=&\frac{%
E_{7\left( 7\right) }}{SO\left( 4,4\right) }  \label{sergio-1} \\
\mathcal{O}_{\mathcal{N}=8,\frac{1}{8}\text{-BPS},p=2,\mathbf{II}} &=&\frac{%
E_{7\left( 7\right) }}{SO\left( 5,3\right) }.  \label{sergio-2}
\end{eqnarray}

In the $\mathcal{N}=2$ exceptional case, the chain of relevant group
branchings reads
\begin{equation}
\mathcal{N}=2,J_{3}^{\mathbb{O}}~\text{nBPS}:E_{7\left( -25\right)
}\longrightarrow E_{6\left( -14\right) }\longrightarrow F_{4\left(
-20\right) }\longrightarrow \left\{
\begin{array}{l}
SO\left( 9\right) \longrightarrow SO\left( 8\right) \\
\mathit{or} \\
SO\left( 8,1\right) \longrightarrow
\begin{array}{l}
SO\left( 8\right) \\
\mathit{or} \\
SO\left( 7,1\right)
\end{array}
\end{array}
\right. \,,
\end{equation}
such that two non-BPS, $\mathcal{N}=2$, $2$-centered charge orbits exist:
\begin{eqnarray}
\mathcal{O}_{\mathcal{N}=2,J_{3}^{\mathbb{O}},\text{nBPS},p=2,\mathbf{I}} &=&%
\frac{E_{7\left( -25\right) }}{SO\left( 8\right) }  \label{first} \\
\mathcal{O}_{\mathcal{N}=2,J_{3}^{\mathbb{O}},\text{nBPS},p=2,\mathbf{II}}
&=&\frac{E_{7\left( -25\right) }}{SO\left( 7,1\right) }.  \label{first-1}
\end{eqnarray}

As it holds for the stabilizer of $\mathcal{O}_{\mathcal{N}=2,J_{3}^{\mathbb{%
O}},\text{BPS},p=2}$ (see Table 1), the Lie algebra $\frak{so}\left(
8\right) $ of the stabilizer of $\mathcal{O}_{\mathcal{N}=2,J_{3}^{\mathbb{O}%
},\text{nBPS},p=2,\mathbf{I}}$ (\ref{first}) is nothing but the Lie algebra $%
\frak{tri}\left( \mathbb{O}\right) $ of the automorphism group $Aut\left(
\mathbf{t}\left( \mathbb{O}\right) \right) $ of the \textit{normed triality}
over the octonionic division algebra $\mathbb{O}$ (see \textit{e.g.} Eq.
(21) of \cite{Baez-O}). It is here worth observing that the Lie algebra $%
\frak{so}\left( 4,4\right) $ of the stabilizer of $\mathcal{O}_{\mathcal{N}%
=8,\frac{1}{8}\text{-BPS},p=2,\mathbf{I}}$ (\ref{sergio-1}) enjoys an
analogous interpretation as the Lie algebra $\frak{tri}\left( \mathbb{O}%
_{s}\right) $ of the automorphism group $Aut\left( \mathbf{t}\left( \mathbb{O%
}_{s}\right) \right) $ of the \textit{normed triality} over the \textit{split%
} form $\mathbb{O}_{s}$ of the octonions. On the other hand, a similar
interpretation seems not to hold for the stabilizer of $\mathcal{O}_{%
\mathcal{N}=8,\frac{1}{8}\text{-BPS},p=2,\mathbf{II}}$ (\ref{sergio-2}) as
well as for the stabilizer of $\mathcal{O}_{\mathcal{N}=2,J_{3}^{\mathbb{O}},%
\text{nBPS},p=2,\mathbf{II}}$ (\ref{first-1}).\medskip\

We expect the $\mathcal{N}=8$ orbits (\ref{sergio-1}) and (\ref{sergio-2}),
as well as the $\mathcal{N}=2$ orbits (\ref{first}) and (\ref{first-1}), to
be defined by different constraints on the four $SL_{h}\left( 2,\mathbb{R}%
\right) \times G_{4}$ invariant polynomials given by Eq. (\ref{four?}); we
leave this interesting issue for further future investigation.\smallskip

Here, we confine ourselves to present parallel results on pseudo-orthogonal
groups, which may shed some light on the whole framework. Let us consider
two vectors $\mathbf{x}$ and $\mathbf{y}$ in a pseudo-Euclidean $\left(
p+q\right) $-dimensional space $E_{p,q}$ with signature $\left( p,q\right) $
and $p>1$, $q>1$. The norm of a vector is defined as, say
\begin{equation}
\mathbf{x}^{2}\equiv x_{1}^{2}+...+x_{p}^{2}-x_{p+1}^{2}-...-x_{p+q}^{2},
\end{equation}
and the scalar product as
\begin{equation}
\mathbf{x}\cdot \mathbf{y}\equiv
x_{1}y_{1}+...+x_{p}y_{p}-x_{p+1}y_{p+1}-...-x_{p+q}y_{p+q}.
\end{equation}
The one-vector orbits (for non-lightlike vectors) are well
\begin{eqnarray}
\mathcal{O}_{p=1,\text{timelike}} &=&\frac{SO\left( p,q\right) }{SO\left(
p-1,q\right) }\text{~if~}\mathbf{x}^{2}>0; \\
\mathcal{O}_{p=1,\text{spacelike}} &=&\frac{SO\left( p,q\right) }{SO\left(
p,q-1\right) }\text{~if~}\mathbf{x}^{2}<0.
\end{eqnarray}
It is intuitively clear that the two-vector orbits do depend on the nature
of the vectors themselves. Let us start and consider two timelike vectors ($%
\mathbf{x}^{2}>0$ and $\mathbf{y}^{2}>0$), whose one-center orbits are
separately given by $\mathcal{O}_{p=1,\text{timelike}}$. It is
straightforward to show that the two-center orbits supporting this
configuration are
\begin{eqnarray}
\frac{SO\left( p,q\right) }{SO\left( p-2,q\right) }\text{~if~}\mathbf{x}^{2}%
\mathbf{y}^{2} &>&\left( \mathbf{x}\cdot \mathbf{y}\right) ^{2};  \label{uno}
\\
\frac{SO\left( p,q\right) }{SO\left( p-1,q-1\right) }\text{~if~}\mathbf{x}%
^{2}\mathbf{y}^{2} &<&\left( \mathbf{x}\cdot \mathbf{y}\right) ^{2}.
\label{due}
\end{eqnarray}
If both vectors are spacelike ($\mathbf{x}^{2}<0$ and $\mathbf{y}^{2}<0$),
the two-center orbits read
\begin{eqnarray}
\frac{SO\left( p,q\right) }{SO\left( p,q-2\right) }\text{~if~}\mathbf{x}^{2}%
\mathbf{y}^{2} &>&\left( \mathbf{x}\cdot \mathbf{y}\right) ^{2};  \label{tre}
\\
\frac{SO\left( p,q\right) }{SO\left( p-1,q-1\right) }\text{~if~}\mathbf{x}%
^{2}\mathbf{y}^{2} &<&\left( \mathbf{x}\cdot \mathbf{y}\right) ^{2}.
\label{quattro}
\end{eqnarray}
Finally, if one vector is timelike and the other one is spacelike (say, $%
\mathbf{x}^{2}>0$ and $\mathbf{y}^{2}<0$), the two-center orbit is unique:
\begin{equation}
\frac{SO\left( p,q\right) }{SO\left( p-1,q-1\right) },  \label{cinque}
\end{equation}
because in this case $\mathbf{x}^{2}\mathbf{y}^{2}<\left( \mathbf{x}\cdot
\mathbf{y}\right) ^{2}$ always holds.

By introducing the $SL_{h}\left( 2,R\right) \times SO\left( p,q\right) $
invariant polynomial (see \cite{Cvetic,Duff-stu} and the last Ref. of \cite
{AM-Refs})
\begin{equation}
\mathbf{I}_{4}\left( \mathbf{x},\mathbf{y}\right) \equiv \mathbf{x}^{2}%
\mathbf{y}^{2}-\left( \mathbf{x}\cdot \mathbf{y}\right) ^{2},
\end{equation}
all orbits (\ref{uno})-(\ref{cinque}) can actually be recognised to
correspond to only three orbits (namely (\ref{uno}), (\ref{tre}), and (\ref
{due})$=$(\ref{quattro})$=$(\ref{cinque})), respectively defined by the $%
\left[ SL_{h}\left( 2,R\right) \times SO\left( p,q\right) \right] $%
-invariant constraints: $\mathbf{I}_{4}>0$ (with $\mathbf{x}^{2}>0$ and $%
\mathbf{y}^{2}>0$); $\mathbf{I}_{4}>0$ (with $\mathbf{x}^{2}<0$ and $\mathbf{%
y}^{2}<0$); $\mathbf{I}_{4}<0$. Note that in the compact case (Euclidean
signature: $q=0$) $\mathbf{I}_{4}>0$ due to the Cauchy-Schwarz triangular
inequality, and the two-vector orbit is unique: $\frac{SO\left( p\right) }{%
SO\left( p-2\right) }$. This is in analogy with the results (obtained in the
complex field) discussed in Section \ref{Counting-Analysis}.

\section*{Acknowledgments}

S. F. and A. M. would like to thank Raymond Stora, Emanuele Orazi and Armen
Yeranyan for useful discussions.

The work of S. F. is supported by the ERC Advanced Grant no. 226455, \textit{%
``Supersymmetry, Quantum Gravity and Gauge Fields''} (\textit{SUPERFIELDS}).


\begin{thebibliography}{99}
\bibitem{DM-1}  F. Denef and G. W. Moore, \textit{Split States, Entropy
Enigmas, Holes and Halos}, \texttt{hep-th/0702146}. F. Denef, D. Gaiotto, A.
Strominger, D. Van den Bleeken and X. Yin, \textit{Black Hole Deconstruction}%
, \texttt{hep-th/0703252}. F. Denef and G. W. Moore, \textit{How many black
holes fit on the head of a pin?}, Gen. Rel. Grav. \textbf{39}, 1539 (2007),
\texttt{arXiv:0705.2564 [hep-th]}.

\bibitem{AM-Refs}  S. Ferrara, R. Kallosh and A. Strominger, $\mathcal{N}%
\mathit{=2}$\textit{\ extremal black holes}, Phys. Rev. \textbf{D52}, 5412
(1995), \texttt{hep-th/9508072}. A. Strominger, \textit{Macroscopic entropy
of }$\mathcal{N}\mathit{=2}$\textit{\ extremal black holes}, Phys. Lett.
\textbf{B383}, 39 (1996), \texttt{hep-th/9602111}. S. Ferrara and R.
Kallosh, \textit{Supersymmetry and attractors}, Phys. Rev. \textbf{D54},
1514 (1996), \texttt{hep-th/9602136}. S. Ferrara and R. Kallosh, \textit{%
Universality of supersymmetric attractors}, Phys. Rev. \textbf{D54}, 1525
(1996), \texttt{hep-th/9603090}.

\bibitem{FGK}  S.~Ferrara, G. W. Gibbons and R. Kallosh, \textit{Black Holes
and Critical Points in Moduli Space}, Nucl. Phys. \textbf{B500}, 75 (1997),
\texttt{hep-th/9702103}.

\bibitem{ADFT-rev}  L. Andrianopoli, R. D'Auria, S. Ferrara and M.
Trigiante, \textit{Extremal black holes in supergravity}, Lect. Notes Phys.
\textbf{737}, 661 (2008), \texttt{hep-th/0611345}.

\bibitem{D-1}  F. Denef, \textit{Supergravity flows and D-brane stability},
JHEP \textbf{0008}, 050 (2000), \texttt{hep-th/0005049}.

\bibitem{D-2}  F. Denef, B. R. Greene and M. Raugas, \textit{Split attractor
flows and the spectrum of BPS D-branes on the quintic}, JHEP \textbf{0105},
012 (2001), \texttt{hep-th/0101135}.

\bibitem{BD}  B. Bates and F. Denef, \textit{Exact Solutions for
Supersymmetric Stationary Black Hole Composites}, \texttt{hep-th/0304094}.

\bibitem{CJ-1}  E. Cremmer and B. Julia, \textit{The }$\mathcal{N}\mathit{=8}
$\textit{\ Supergravity Theory. 1. The Lagrangian}, Phys. Lett. \textbf{B80}%
, 48 (1978). E. Cremmer and B. Julia, \textit{The }$\mathit{SO(8)}$\textit{\
Supergravity}, Nucl. Phys. \textbf{B159}, 141 (1979).

\bibitem{HT-1}  C. Hull and P. K. Townsend, \textit{Unity of Superstring
Dualities}, Nucl. Phys. \textbf{B438}, 109 (1995), \texttt{hep-th/9410167}.

\bibitem{David}  J. R. David, \textit{On walls of marginal stability in }$%
\mathcal{N}\mathit{=2}$\textit{\ string theories}, JHEP \textbf{0908}, 054
(2009), \texttt{arXiv:0905.4115 [hep-th]}.

\bibitem{MS-FMO-1}  S. Ferrara, A. Marrani and E. Orazi, \textit{Split
Attractor Flow in }$\mathcal{N}\mathit{=2}$\textit{\ Minimally Coupled
Supergravity}, \texttt{arXiv:1010.2280 [hep-th]}.

\bibitem{FMOSY-1}  S. Ferrara, A. Marrani, E. Orazi, R. Stora and A.
Yeranyan, \textit{Two-Center Black Holes Duality-Invariants for }$\mathit{stu%
}$\textit{\ Model and its lower-rank Descendants}, \texttt{arXiv:1011.5864
[hep-th]}.

\bibitem{Luciani}  J. F. Luciani, \textit{Coupling of O(2) Supergravity with
Several Vector Multiplets}, Nucl. Phys. \textbf{B132}, 325 (1978).

\bibitem{GST}  M. G\"{u}naydin, G. Sierra and P. K. Townsend, \textit{%
Exceptional Supergravity Theories and the Magic Square}, Phys. Lett. \textbf{%
B133}, 72 (1983). M. G\"{u}naydin, G. Sierra and P. K. Townsend, \textit{The
Geometry of }$\mathcal{N}\mathit{=2}$\textit{\ Maxwell-Einstein Supergravity
and Jordan Algebras}, Nucl. Phys. \textbf{B242}, 244 (1984).

\bibitem{dWVVP}  B. de Wit, F. Vanderseypen and A. Van Proeyen, \textit{%
Symmetry Structures of Special Geometries}, Nucl. Phys. \textbf{B400}, 463
(1993), \texttt{hep-th/9210068}.

\bibitem{BFGM1}  S. Bellucci, S. Ferrara, M. G\"{u}naydin and A. Marrani,
\textit{Charge orbits of symmetric special geometries and attractors}, Int.
J. Mod. Phys. \textbf{A21}, 5043 (2006), \texttt{hep-th/0606209}.

\bibitem{GLS-2}  E. G. Gimon, F. Larsen and J. Simon, \textit{Constituent
Model of Extremal non-BPS Black Holes}, JHEP \textbf{0907}, 052 (2009),
\texttt{arXiv:0903.0719 [hep-th]}.

\bibitem{CS}  A. Castro and J. Simon, \textit{Deconstructing the }$\mathit{D0%
}$\textit{-}$\mathit{D6}$\textit{\ system}, JHEP \textbf{0905}, 078 (2009),
\texttt{arXiv:0903.5523 [hep-th]}.

\bibitem{Gunaydin-Lects}  M. G\"{u}naydin, \textit{Lectures on Spectrum
Generating Symmetries and }$\mathit{U}$\textit{-Duality in Supergravity,
Extremal Black Holes, Quantum Attractors and Harmonic Superspace}, \texttt{%
arXiv:0908.0374 [hep-th]}.

\bibitem{Baez-O}  J. C. Baez, \textit{The Octonions}, Bull. Am. Math. Soc.
\textbf{39}, 145 (2001), \texttt{math/0105155}.

\bibitem{FGimK}  S. Ferrara, E. G. Gimon and R. Kallosh, \textit{Magic
Supergravities,} $\mathcal{N}\mathit{=8}$ \textit{and Black Hole Composites}%
, Phys. Rev. \textbf{D74}, 125018 (2006), \texttt{hep-th/0606211}.

\bibitem{ADFGT}  L. Andrianopoli, R. D'Auria, S. Ferrara, P. A. Grassi and
M. Trigiante, \textit{Exceptional} $\mathcal{N}\mathit{=6}$ \textit{and} $%
\mathcal{N}\mathit{=2}$ $AdS_{4}$ \textit{Supergravity, and Zero-Center
Modules}, JHEP \textbf{0904}, 074 (2009), \texttt{arXiv:0810.1214 [hep-th]}.

\bibitem{ADF}  L. Andrianopoli, R. D'Auria and S. Ferrara, $\mathit{U}$%
\textit{-invariants, black hole entropy and fixed scalars}, Phys. Lett.
\textbf{B403}, 12 (1997), \texttt{hep-th/9703156}.

\bibitem{RS}  D. Roest and H. Samtleben, \textit{Twin Supergravities},
Class. Quant. Grav. \textbf{26}, 155001 (2009), \texttt{arXiv:0904.1344
[hep-th]}.

\bibitem{Slansky}  R. Slansky, \textit{Group Theory for Unified Model
Building}, Phys. Rept. \textbf{79}, 1 (1981).

\bibitem{deWit:2002vt}  B.~de Wit, H.~Samtleben and M.~Trigiante, \textit{On
Lagrangians and gaugings of maximal supergravities}, Nucl.\ Phys.\ \textbf{%
B655} (2003) 93, \texttt{hep-th/0212239}.

\bibitem{Exc-Reds}  A. Marrani, E. Orazi and F. Riccioni, \textit{%
Exceptional Reductions}, \texttt{arXiv:1012.5797v1 [hep-th]}.

\bibitem{G-1}  D. Gaiotto, W. W. Li and M. Padi, \textit{Non-Supersymmetric
Attractor Flow in Symmetric Spaces}, JHEP \textbf{0712}, 093 (2007), \texttt{%
arXiv:0710.1638 [hep-th]}.

\bibitem{Duff-Freud}  L. Borsten, D. Dahanayake, M. J. Duff and W. Rubens,
\textit{Black Holes Admitting a Freudenthal Dual}, Phys. Rev. \textbf{D80},
026003 (2009), \texttt{arXiv:0903.5517 [hep-th]}.

\bibitem{MS-FM-1}  S. Ferrara and A. Marrani, \textit{Matrix Norms, BPS
Bounds and Marginal Stability in }$\mathcal{N}\mathit{=8}$\textit{\
Supergravity}, JHEP (2010), in press, \texttt{arXiv:1009.3251 [hep-th]}.

\bibitem{Bossard}  G. Bossard, $\mathit{1/8}$\textit{\ BPS black hole
composites}, \texttt{arXiv:1001.3157 [hep-th]}.

\bibitem{FG1}  S. Ferrara and M. G\"{u}naydin, \textit{Orbits of Exceptional
Groups, Duality and BPS States in String Theory}, Int. J. Mod. Phys. \textbf{%
A13}, 2075 (1998), \texttt{hep-th/9708025}.

\bibitem{Cvetic}  M. Cvetic and D. Youm, \textit{Dyonic BPS saturated black
holes of heterotic string on a six torus}, Phys. Rev. \textbf{D53}, 584
(1996), \texttt{hep-th/9507090}. M. Cvetic and A. A. Tseytlin, \textit{%
General class of BPS saturated dyonic black holes as exact superstring
solutions}, Phys. Lett. \textbf{B366}, 95 (1996), \texttt{hep-th/9510097}.
M. Cvetic and A. A. Tseytlin, \textit{Solitonic strings and BPS saturated
dyonic black holes}, Phys. Rev. \textbf{D53}, 5619 (1996); Erratum-ibid.
\textbf{D55}, 3907 (1997), \texttt{hep-th/9512031}.

\bibitem{Duff-stu}  M. J. Duff, J. T. Liu and J. Rahmfeld, \textit{%
Four-dimensional string/string/string triality}, Nucl. Phys. \textbf{B459},
125 (1996), \texttt{hep-th/9508094}.
\end{thebibliography}
\end{document}